\documentstyle[epsfig,12pt,twoside]{fearticle}

\newcommand{\beq}{\vspace{-3mm}\begin{equation}}
\newcommand{\eeq}{\end{equation}\vspace{-4mm}}
\newcommand{\bea}{\vspace{-3mm}\begin{array}}
\newcommand{\ena}{\end{array}\vspace{-1mm}}

\newcommand{\authors}{JORGE \'I\~NIGUEZ {\sl et al.}}
\newcommand{\atit}{FIRST-PRINCIPLES STUDY OF THE STRUCTURAL ...}

\newcommand{\rheader}[1]{\markright{\protect\rule[-5pt]{16cm}{.5pt}\kern-16cm
    {\normalsize #1}}}
\newcommand{\lheader}[1]{\markright{\protect\rule[-5pt]{16cm}{.5pt}\kern-16cm
                         {\normalsize #1}}}
\newcommand{\nheader}[1]{\markright{\protect\rule[-5pt]{16cm}{.5pt}\kern-16cm
    {\cysc #1}}}

\title{FIRST-PRINCIPLES STUDY OF THE STRUCTURAL INSTABILITIES IN
HEXAGONAL BARIUM TITANATE: COUPLING BETWEEN THE SOFT OPTICAL AND THE
ACOUSTIC MODES}

\author{JORGE \'I\~NIGUEZ$^*$, ALBERTO
GARC\'IA$^*$\\ and J. M. P\'EREZ-MATO$^{\dagger}$\\
\vskip	1em
	Departamentos de $^*$F\'{\i}sica Aplicada II\\
	y $^\dagger$F\'{\i}sica de la Materia Condensada.\\
	Universidad del Pa\'{\i}s Vasco, Apdo 644, 48080 Bilbao, Spain
}

\begin{document}
\markboth{\authors}{\atit}

\thispagestyle{empty} 
\maketitle

\begin{abstract}
Hexagonal BaTiO$_3$ undergoes a structural phase transition to an
orthorhombic $C222_1$ phase at $T_0 = 222$ K. The transition is driven
by a soft optical mode with $E_{2u}$ symmetry whose couplings force
the appearance of a spontaneous $E_{2g}$ strain (improper ferroelastic
character). Staying within the same $E_{2u}$ subspace, the system
could in principle settle into a second (not observed) orthorhombic
phase ($Cmc2_1$). We have carried out a first-principles investigation
of these questions, studying the structure of the existing $C222_1$
and the {\sl virtual} $Cmc2_1$ phases, and describing the spontaneous
$E_{2g}$ strain in accord with the experimental observations. In
addition, we show that the occurrence of $C222_1$ instead of $Cmc2_1$
cannot be explained by the $E_{2u}$ soft modes themselves and,
therefore, must be related to their couplings with secondary order
parameters. A more detailed analysis proves that the $E_{2g}$ strains
do not account for the experimental preference.
\end{abstract}


\section{INTRODUCTION}

Barium titanate has two structural polymorphs: the well known
perovskite type (p-BT) and its hexagonal modification (h-BT) with 30
atoms per unit cell. h-BT Undergoes two zone-center structural phase
transitions: at $T_0=222$ K from the hexagonal $P6_3/mmc$ phase (I) to
an orthorhombic $C222_1$ (II), and at $T_C=74$ K to a monoclinic $P2_1$
phase (III).\cite{Ak94} The transition at $T_0$ presents two
specially interesting features:
\begin{itemize}
\item It is driven by a ``silent'' soft optical mode of $E_{2u}$
symmetry,\cite{IH88} coupled quadratically to the $E_{2g}$ acoustic
modes (strains) $\eta_1 - \eta_2$ and $\eta_6$, which spontaneously
acquire non-zero values at the transition.\cite{HS89} In the usual
terminology, the transition is then improper ferroelastic.  The
coupling of the $E_{2g}$ modes with the soft optical modes has been
studied spectroscopically.\cite{YW95,YW96}

\item Depending on the direction taken by the bidimensional $E_{2u}$
order parameter, the system can find itself in structures of three
different symmetries: the experimentally observed $C222_1$ and $P2_1$,
as well as a second orthorhombic $Cmc2_1$ phase (II$'$). The
preference of II over II$'$ must ultimately
be dictated by the energetics of the $E_{2u}$ instability and by its
couplings with secondary order parameters.

\end{itemize}
In this paper we present a summary of the results of our investigation
into the $E_{2u}$ optical instability in h-BT, including its couplings
with the spontaneous $E_{2g}$ strains, and into the issue of the
relative stability of the two accessible orthorhombic phases, using
first-principles density-functional methods. To our knowledge, this is
the first {\sl ab initio} study of hexagonal BaTiO$_3$.

\section{TECHNICAL DETAILS}

For our calculations we have used density-functional theory within the
pseudopotential approach and the local-density
approximation.\cite{Pi89} We have employed Vanderbilt's ultrasoft
pseudopotentials,\cite{Va90} which are ideally suited to deal with
first-row elements and transition metals. The electron wave functions
were expanded in a plane wave basis with an energy cut-off of 25
Rydberg, and the Brillouin zone sums were calculated by a
$3\times3\times(2(+0.5))$ Monkhorst-Pack special k-point
mesh.\cite{MP76} We checked the validity of these approximations by a
convergence analysis. This first-principles framework has been
successfully used for the investigation of structural properties in
perovskite oxides.\cite{Va97}

\section{RESULTS AND DISCUSSION}

In order to proceed, the reference hexagonal structure must be
determined. In I the occupied Wyckoff positions are fixed by symmetry
up to five free parameters. We used the systematic procedure of
Ref.~\citen{VL84} to relax these five $A_{1g}$ optical modes so that
the forces on atoms were below $10^{-4}$ eV$\cdot$bohr$^{-1}$. Our
results for the fractional atomic coordinates agree with
experiment\cite{AG94} to within $0.6\%$. Due to the high computational
cost of a complete analysis, we used the experimental values for the
unit cell parameters ($a=10.77$ bohr and $c/a=2.456$).

In what follows we work with a symmetry-adapted Taylor expansion of
the potential energy of the system around the reference structure, as
a function of the above mentioned relevant variables: the $E_{2u}$
soft optical and the $E_{2g}$ acoustic modes. It is convenient to
decompose this expansion into:

\beq E \,- \,E_{\rm I} \;= \;E_{\rm optical} \,+ \,E_{\rm strain} \,+
\,E_{\rm coupling},\label{E}\eeq

where $E_{\rm I}$ is the energy of the reference structure I.

As a pre-requisite for the study of $E_{\rm optical}$, we needed to
check that the {\sl ab initio} methods predict the experimentally
observed instability. We solved the associated eigenvalue problem
(there are seven zone center $E_{2u}$ symmetry modes) and indeed found
only one instability with spring constant $-0.743$
eV$\cdot$bohr$^{-2}$. (We consider the Fourier transform of the so
called force-constant-matrix and not the dynamical matrix.) The
corresponding (degenerate) eigenvectors determine the unstable
two-dimensional subspace. Using as coordinates the respective
amplitudes in polar form ($Q_1 =: r cos\varphi$ and $Q_2 =: r
sin\varphi$), we can write:

\beq E_{\rm optical} \;= \;\frac{1}{2} A_2 r^2 \,+ \,\frac{1}{4} A_4
r^4 \,+ \,\frac{1}{6} \left[A_6 + A_6' cos (6\varphi)\right] r^6,
\label{E-opt}\eeq

where $\varphi = \{0, \pm \pi /3, \pm 2\pi /3, \pi\}$ and $\varphi =
\{\pm \pi /6, \pm \pi / 2, \pm 5\pi /6\}$ correspond to II and II$'$
respectively, while for a general $\varphi$ the system is in III.
Eq.(\ref{E-opt}) shows that the sign of $A_6'$ determines the relative
stability of the orthorhombic minima (with $A_2<0$).\footnote[3]{
For the energetic analysis of the monoclinic phases, higher order
terms should be considered.}

\begin{figure}[h!]
\begin{center}
\epsfig{file=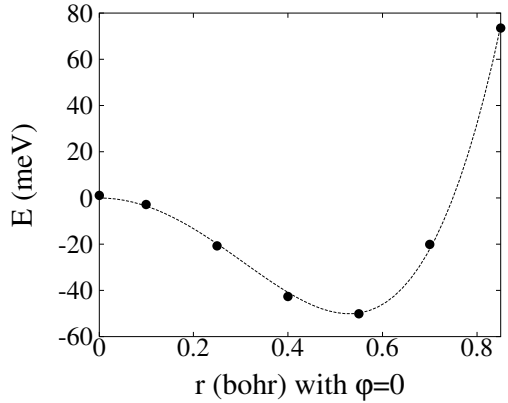,width=5cm,angle=-90}
\end{center}
\caption{Potential energy well of the $E_{2u}$ instability of I, in
the $\varphi=0$ direction (corresponding to II). The {\sl ab initio}
data is fitted to a quartic and $E_{\rm I}$ is set to zero.}
\label{fig-opt}
\end{figure}

Figure (\ref{fig-opt}) shows $E_{\rm optical}$ as a function of
$r$ with $\varphi=0$, the case of II. (The data for
II$'$ are identical at the scale of the figure). In order to model
these data, two facts must be taken into account. First, quadratic
($A_2$) and quartic ($A_4$) terms are symmetry-forced to be the same
in both phases. Second, the fitted $A_2$ must be in good accord with
the previously obtained eigenvalue.  We determined: $A_2 = -0.72\pm
0.02$ eV$\cdot$bohr$^{-2}$ and $A_4 = 2.56\pm 0.07$
eV$\cdot$bohr$^{-4}$, with the energy minima in $|r_{min}| = 0.52\pm
0.01$ bohr ($E_{min} = -50\pm 4$ meV).  We could obtain no convincing
fitting for the $O(r^6)$ term and thus concluded that the quartic
approximation constitutes the best model to describe the $E_{2u}$
instabilities, i.e., the soft modes themselves cannot account for the
differences between the II and II$'$ minima, so their couplings with
secondary order parameters must be responsible for the experimental
observation of II.

This calculation constitutes a prediction of the structural parameters
of the orthorhombic $C222_1$ phase, which to our knowledge has not yet
been determined experimentally. (Further details about the structure will
be published elsewhere.)

Turning to the pure elastic energy term in
Eq.(\ref{E-opt}), it is convenient to use polar coordinates also for
the $E_{2g}$ strains ($\eta_1 - \eta_2$ =: $\rho cos\phi$ and $\eta_6$
=: $\rho sin\phi$), and write:

\beq E_{\rm strain} \;= \;C_{2}\rho^2 \,+ \,C_{3}cos (3\phi)
\rho^3.\label{E-str}\eeq

For a general $\phi$ the system is distorted to III (due to $\eta_6$),
while for $\phi = \{0, \pi\}$ it goes to $Cmcm$, which is a supergroup
of both II and II$'$, so $\eta_1 - \eta_2$ is the only secondary order
parameter we take into account. We studied the response of the system
to this strain with the optical phonon amplitudes set to zero (and
imposing $\eta_1 + \eta_2 = 0$, so that only deformations with
$E_{2g}$ symmetry are considered), and concluded that the best model
reduces to $C_{2}=130\pm 2$ eV. (The $O(\rho^3)$ term cannot be
reliably fitted.)

Finally, we examined the coupling between the $E_{2u}$ soft optical and
the $E_{2g}$ acoustic degrees of freedom, considering:

\beq \bea{ll} E_{\rm coupling} = & E^{(1)}_{\rm coupling} 
+ E^{(2)}_{\rm coupling} \\
E^{(1)}_{\rm coupling} = & \alpha_{12} cos (2\varphi + \phi)\rho r^2 \\ 
E^{(2)}_{\rm coupling} =& \left\{\alpha_{14} cos(2\varphi + \phi) + \beta_{14}
cos(4\varphi + \phi)\right\} \rho r^4 \\&+\left\{ \alpha_{22} +
\beta_{22} cos\left[2(\varphi - \phi)\right]\right\} \rho^2 r^2
\\&+\left\{ \alpha_{24} + \beta_{24} cos\left[2(\varphi -
\phi)\right] + \gamma_{24} cos\left[2(2\varphi - \phi)\right]\right\}
\rho^2 r^4, \label{E-cou}\ena\eeq

\vspace{.5cm}

\begin{figure}[h!]
\begin{center}
\epsfig{file=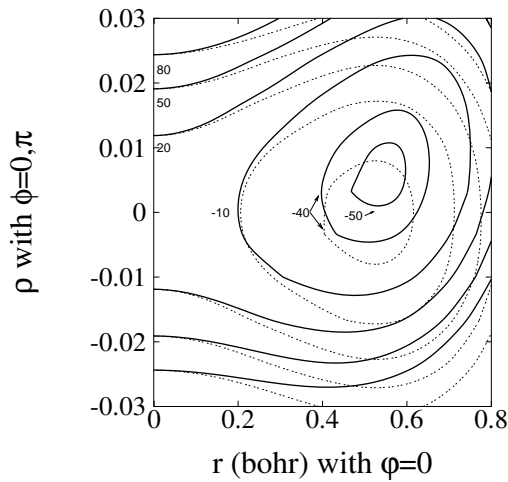,width=6cm
,angle=-90}
\end{center}
\caption{$(r,\rho)$ Energy map of II. We show the equi-energy lines of
a model in which only the $A_2$, $A_4$ and $C_{2}$ terms (with the
values of the text) are present (dashed) and those corresponding to
the fit in table (\ref{table-cou}) (solid). Energies are given in meV
and $E_{\rm I}$ is set to zero.}
\label{fig-cou}
\end{figure}

The first term $E^{(1)}_{\rm coupling}$ is the simplest form of
coupling usually considered in the literature,\cite{IH92,IT89} and
indeed it provides us with a very good fit of the ab-initio energies
in the presence of $E_{2g}$ distortions (for both the II and II$'$
phases). Figure (\ref{fig-cou}) shows the main result of including
this minimal strain coupling into the energy analysis: the minima
shift their positions and get deeper. As table (\ref{table-cou})
shows, the configurations of minimum energy are located at $|r_{min}|
= 0.55\pm 0.01$ bohr, and the associated spontaneous strain is
$\rho_{min} = (5.6\pm 0.2)\times 10^{-3}$ with $\phi = 0$ and $\pi$
for II and II$'$ respectively. We found no experimental measurement of
the $\alpha_{12}$ parameter; nevertheless, its sign is correct as
compared with results in Ref.~\citen{YY91}. The location of the wells
provides us with a more complete (this time including strain)
prediction of the structure of the orthorhombic phase.

\begin{table}[h!]
\begin{center}
\begin{tabular}{|c|ccc|}
\hline

            & II       & II$'$ & II \& II$'$  \\

\hline

$C_{2}$ (eV)   & $134\pm 2$ & $133\pm 2$   & $133\pm 2$ \\

$\alpha_{12}$ (eV$\cdot$bohr$^{-2}$) &$-5.0\pm 0.1$ & $-4.8\pm 0.1$ &
$-4.9\pm 0.1$ \\

\hline

$\rho_{min}$ ($10^{-3}$) & $5.6\pm 0.2$ & $-5.4\pm 0.2$ & $(\pm)5.6\pm
0.2$ \\

$|r_{min}|$ (bohr) & $0.55\pm 0.01$ & $0.55\pm 0.01$ & $0.55\pm
0.01$\\

$E_{min}$ (meV) & $-54\pm 4$ & $-54\pm 4$ & $-54\pm 4$\\

\hline
\end{tabular}
\end{center}
\caption{Best fits for Eq.(\ref{E}) when only a minimal coupling
($\alpha_{12}$ term) is considered in the model. We present the fits
corresponding to II and II$'$ separately (so that the fulfillment of
symmetry restrictions can be observed), plus a final result (II \&
II$'$) including aggregate data for the two phases. In all cases,
parameters $A_2$ and $A_4$ were fixed to the values in the text. The
quality of the fits is highly sensitive to small variations in
$C_{2}$, so we left it free (the obtained results are within the error
of our initial estimation).}
\label{table-cou}

\end{table}

However, as the well depths are within this minimal model identical
for both II and II$'$, we still have found no reason for Nature's
preference for the first. To further analyze this issue, we included
$E^{(2)}_{\rm coupling}$ in Eq.(\ref{E-cou}) in our fitting procedure
and found that only the $O(\rho^2 r^4)$ terms improve the model. We
got $\alpha_{24}+\beta_{24}+ \gamma_{24}=7.3\pm 0.2$
eV$\cdot$bohr$^{-4}$ for II and
$\alpha_{24}-\beta_{24}+\gamma_{24}=7.9\pm 0.2$ eV$\cdot$bohr$^{-4}$
for II$'$, but the difference is so small that it produces no definite
distinction between the two minima. Thus, it can be concluded that
neither the $E_{2u}$ instability itself nor its coupling with $E_{2g}$
strains justify the existence of II instead of II$'$.

\section{CONCLUSIONS}

We have presented a first-principles study on the stability of
hexagonal barium titanate when distorted by the zone center $E_{2u}$
soft optical and the $E_{2g}$ acoustic modes.

We have shown that first-principles calculations are able to predict
the experimentally observed zone center $E_{2u}$ instability and to
define the corresponding unstable subspace of distortions. We have
examined the anharmonicity of the soft modes and obtained a prediction
of the structure of the observed $C222_1$ and of the {\sl virtual}
$Cmc2_1$ phases, including the appearance of a spontaneous $E_{2g}$
deformation which is coupled to the soft mode.  Moreover, we find that
the two accessible orthorhombic minima are not definitely
distinguished by the $E_{2u}$ instability itself, which points to the
importance of secondary order parameters to decide this
issue. However, the effect of the coupled $E_{2g}$ strains is not
large enough to explain the experimental observation of $C222_1$
instead of $Cmc2_1$, which must then be due to the couplings of the
soft mode with other secondary order parameters.

\section{ACKNOWLEDGEMENTS}

This work was supported in part by the UPV research grants
060.310-EA149/95 and 063.310-G19/98, and by the Spanish Ministry of
Education grant PB97-0598. J.I. acknowledges fellowship support from
the Basque regional government and thanks Inmaculada Peral and Agustin
V\'algoma for their kind encouragement.

\end{document}